\documentclass[preprint,showpacs,preprintnumbers,amsmath,amssymb]{revtex4}

\usepackage{graphicx}
\usepackage{dcolumn}
\usepackage{bm}
\usepackage{algorithm}
\usepackage[noend]{algpseudocode}
\usepackage{hyperref}

\DeclareMathOperator*{\argmin}{arg\,min}

\begin{document}

\preprint{APS/123-QED}

\title{Computing the statistical significance of optimized communities in networks}

\author{John Palowitch}
 \email{palowitch@google.com}
\affiliation{%
Google Research\\
1600 Ampitheatre Parkway\\
Mountain View, CA 94043
}%

\date{\today}

\begin{abstract}
It is often of interest to find communities in network data for unsupervised learning, feature discovery, anomaly detection, or scientific study. The vast majority of community detection methods proceed via optimization of a quality function, which is possible even on random networks without communities. Therefore there is usually not an easy way to tell if a community is ``significant'', in this context meaning more internally connected than would be expected under a random graph model without communities. This paper generalizes existing null models for this purpose to bipartite graphs, and introduces a new significance scoring algorithm called Fast Optimized Community Significance (FOCS) that is highly scalable and agnostic to the type of graph. Furthermore, compared with existing methods on unipartite graphs, FOCS is more numerically stable and better balances the trade-off between detection power and false positives.
\end{abstract}

\maketitle

\section{\label{s:intro}Introduction}

Many natural systems can be modeled as a network, with network nodes representing entities and network edges representing links or relationships between those entities. As such, a wide variety of network models and graph algorithms have been developed, generalized, and improved over many decades, forming the field of network science and the study of complex networks \citep{boccaletti2006complex}. A sub-field of network science is focused on methodology for and applications of ``community'' detection. Defined loosely, a community is a subset of nodes in a network that are more connected to each other than they are to other nodes. There are many distinct, precise definitions of a community, with utilities that vary by application \citep{fortunato2016community}. In practice, the purpose of community detection is to discover dynamics or features of the networked system that were not known in advance. Community detection has been profitably applied to naturally arising networks in diverse fields like machine learning, social science, and computational biology \citep{fortunato2010community}.

Usually, the aim of community detection is to find communities in a network that are optimal with respect to some quality function or search procedure. Often, a \emph{partition} of the network is the object being optimized with the quality function. Arguably the most commonly used and studied quality function for partition optimization is modularity, which is the sum of the first-order deviations of each community's internal edge count from a random graph null model \citep{newman2004finding}. Other community detection methods aim to find a \emph{collection} of communities, where the requirement that communities be disjoint and exhaustive is relaxed (partitions are also collections). In some approaches, collections of communities are found by optimizing communities one-by-one, according to a community-level quality function \citep{zhao2011community, lancichinetti2011finding, wilson2017community}.

Hundreds of distinct community detection methods have been introduced in recent decades. Despite this, relatively few articles discuss issues of statistical significance related to community detection. In particular, there is often no immediate way to determine if the communities returned by a community detection algorithm are of higher ``quality'' than would be expected (on average) if the algorithm were run repeatedly on a random graph model without true communities. When significance is discussed or addressed, it is usually with reference to the overall partition, rather than individual communities \citep[e.g.][]{zhang2014scalable, peixoto2015model}.

This paper introduces a method called Fast Optimized Community Significance (FOCS) for scoring the statistical significance of individual communities that, importantly, have been optimized by a separate method. As discussed in Section \ref{ss:existing}, there are several existing methods that score optimized communities. This paper makes two advancements in this area:
\begin{enumerate}
\item Null models for scoring optimized communities are made explicit and newly generalized to bipartite graphs.
\item A new method (FOCS) is introduced which enjoys some benefits over existing methods:
{
\begin{itemize}
\item A core algorithm that is transparent, easy to implement, and general enough to apply to either unipartite or bipartite graphs without modification.
\item Higher numerical stability and 10-100x faster runtimes
\item Conservative scores on optimized communities from null networks, while maintaining comparable or dominant detection power on true communities.
\end{itemize}
}
\end{enumerate}

In this paper, a network is denoted by $G:=(V, A)$, where $V$ is a set of vertices and $A$ is an adjacency matrix. Let $n:=|V|$. For $u,v\in V$, the entry $A(u,v)$ is equal to the number of edges between nodes $u$ to $v$. Unless explicitly stated otherwise, all networks considered will be undirected, so that $A(u,v) = A(v, u)$. In the following sections, denote the degree of $u\in V$ by $d_u := \sum_{v\in V} A(u,v)$. Let $C\subseteq V$ denote any node subset. With a slight abuse of notation, let $d_C:=\sum_{u\in C}d(u)$ be the total degree of a subset. Analogously, $d_u(C) := \sum_{v\in C} A(u,v)$, and $d_C(C'):=\sum_{u\in C}d_u(C')$, where $C' := V\setminus C$. In general, the notation $d_{a}(B)$ can be read and understood as ``the degree of $a$ in set $B$''. Note that for undirected networks, $d_C(C') = d_{C'}(C)$ for any $C\subseteq V$.

\subsection{Existing work} \label{ss:existing}
Currently there are several methods for scoring optimized communities by statistical significance. In one recent publication, a simulation-based method called the QS-Test was proposed \citep{kojaku2018generalised}. The QS-Test generates 500 independent configuration-model networks, each with a degree distribution matching the observed network. (The repetition count 500 is the default value for the method, and can be changed.) On each network, a community detection algorithm is run, and a kernel density estimator is applied to the resulting sets of quality functions. This provides a null distribution against which to compare observed values of the quality function.

The QS-Test approach has many desirable features. First, it is general, in that it can be applied with any quality function and any community detection algorithm. Furthermore, it is (at least in principle) evaluating community significance against a direct estimate of its quality function's null distribution. However, the approach also has drawbacks. First, unless given an unlimited number of machines, it is not scalable, as it requires hundreds or thousands of simulated graphs with the same number of nodes and edges as the observed network. It also requires the community detection algorithm of choice to be run on each of those networks. Even if a separate machine was available for each simulation, the collection of the results and subsequent density estimation procedure would be cumbersome in an online data setting.

An older approach introduced in \cite{lancichinetti2010statistical} uses an analytical approximation to compute the statistical significance of a community. This approach was referenced in \cite{kojaku2018generalised} and included in that paper's simulation study. The authors of \cite{lancichinetti2010statistical} begin with a conditional configuration model which fixes the number of internal edge counts of the community of interest. Under this model, the edge count $d_u(C)$ of any external node $u\notin C$ follows a certain hypergeometric distribution. The authors reason that, if the community is a false positive, the in-degree of its \emph{worst} node should be distributed as the maximum hypergeometric order statistic of the external nodes. They derive a basic score from this observation, and then propose a modified version of the score for an \emph{optimized} community. The particulars of this method will be discussed further in Section \ref{s:method}, as the FOCS approach has a similar foundation.

Building upon their score based on a community's worst node, the authors then propose to test nodes up to the $k$-th worst node in the community. They show through empirical studies that this ``B-Score'' (for ``border'' score) is more powerful while remaining conservative on false-positive communities. The strengths of the B-Score approach over the QS-Test is that it is analytical and thus faster to compute. A drawback of the approach is that it contains more approximations to the null distribution than the QS-Test, and does not have the notion of effect-size or quality score which is inherent to that method.

\section{The FOCS algorithm}\label{s:method}
The methodology introduced in this paper is based on a conditional configuration model, similar to (but not identical to) that introduced in \cite{lancichinetti2010statistical}. The null model and new methodology used in this paper are described first for unipartite graphs; a novel generalization to bipartite graphs is presented in Section \ref{ss:bipartite}. Given a community of interest $C$, the focus of the core FOCS algorithm is on the edge distribution of external nodes $u\in C'$, under a random graph null model.  The null model breaks edges coming out of $C$, and all edges internal to $C'$, and randomly re-assigns the edges of $u$ without replacement. Under this model, the degree of $u$ in $C$ has a hypergeometric pmf:
\begin{equation}\label{eq:basic-null}
P(d_u(C) = x) \propto \frac{{d_C(C')\choose x}{d_{C'}(C')\choose d_u - x}}{{d_{C'}\choose d_u}}.
\end{equation}
If $C$ is optimized, the least-connected or ``worst'' in-community node $w\in C$ should be at the maximum quantile of $P$, among external nodes. Explicitly, define $g_P(u, C):=P(d_u(C)\leq \tilde{d}_u(C))$, where $\tilde{d}_w(C)$ is the random version of $d_w(C)$ with respect to $P$. Define the worst node by $w:= \argmin_{u\in C}g_P(u, C)$. Then, among statistics $\{d_v(C) : v\in\{w\}\cup C'\}$, the random variable $d_w(C)$ should be treated as the maximum order statistic. Thus it is possible to test the significance of $C$ by comparing the $g_P(u, C)$ to the distribution of the minimum of $|C'| + 1$ uniform random variables from the unit line. Writing as $F_m^{(1)}$ the cumulative distribution function of the minimum of $m$ uniform order statistics, the significance score on which FOCS is based is therefore defined

\begin{equation}\label{eq:g-score}
f(C) = F_{|C'| + 1}^{(1)}(g_P(w, C))
\end{equation}
where $F^{(1)}_m$ is CDF of the minimum order statistic from $m$ unit line uniform random variables. The score $f(C)$ has the standard interpretation given to traditional p-values - a low value of $f(C)$ implies that the connectivity observed in $C$ is unlikely to have arisen in a random (community-less) network. 

The idea of using the worst node of a community to test optimized communities was introduced in \cite{lancichinetti2010statistical}. However, those authors proposed adjusted hypergeometric parameters that account for perfect community optimization, which is more fully described in their publication. The approach in the present paper uses the observation that in practice, communities are rarely perfectly optimized. In fact, exact modularity optimization is exponentially complex and computationally infeasible on networks with any more than a few hundred nodes \citep{brandes2007finding}. Furthermore, the modularity maximization surface is glassy, with many local optima extremely close to the true maximum \citep{good2010performance}. This suggests that for a locally optimized, yet truly false positive, community, the distribution of worst nodes can be adequately described by the simpler procedure outlined above. Note that the null model described is well-defined for node pairs with multiple edges.

There may be multiple nodes in an optimized community that are spurious, in the sense that moving them to another community would not significantly change the quality score of the overall partition \cite{good2010performance}. Therefore, instead of a single worst node, a ``worst set'' of nodes may be a more robust test subject for determining significance. To test a worst set of nodes, the FOCS method computes $f(C)$, removes the worst node, re-computes $f$, and so-on until a given proportion $p$ nodes are tested. The pseudocode for FOCS is given in Algorithm \ref{algo:socs}.\\
\begin{algorithm}[H]
\caption{FOCS}\label{algo:socs}
\begin{algorithmic}[1]
\State Given community $C\subseteq V$ and proportion $p\in(0,1]$.
\State Size of test set $k \gets \lceil p|C|\rceil$.
\State Minimum score $m\gets 1$.
\While{$k > 0$}
  \State $w\gets$ worst node in $C$.
  \State $s\gets f(C)$.
  \If{$s < m$}
  	\State $m\gets s$.
  \EndIf
  \State $C \gets C\setminus\{w\}$.
  \State $k\gets k-1$.
\EndWhile\label{testwhile}
\State\textbf{return} $m$.
\end{algorithmic}
\end{algorithm}

In practice, to resolve computational issues arising from lack of continuity in $f$, the cumulative probabilities given by Equation \ref{eq:basic-null} are sampled around their observed values, and the median FOCS score arising from these samples is used. Also, note that the test set proportion $p$ is a free parameter in the method. Setting $p < 0.5$ is the safest, as testing the ``best'' or most interior nodes of an optimized community may lead to spuriously low values of $f$, even under the null, since the community has been optimized. In our simulations and real data applications, a globally-applied setting of $p=0.25$ appears to perform well.

The FOCS algorithm has multiple practical benefits. First, it is simple to implement and fast to compute. Second, testing multiple worst-nodes is beneficial when there are ground-truth communities in the network. As mentioned above, modularity optimization is necessarily local, and thus even real communities may be contaminated with noise nodes. Using FOCS helps to bypass noise nodes in a real community, increasing detection power.

\subsection{Extension to bipartite and directed networks}\label{ss:bipartite}
\begin{figure}[!htb]
\centering
\includegraphics[scale=0.3]{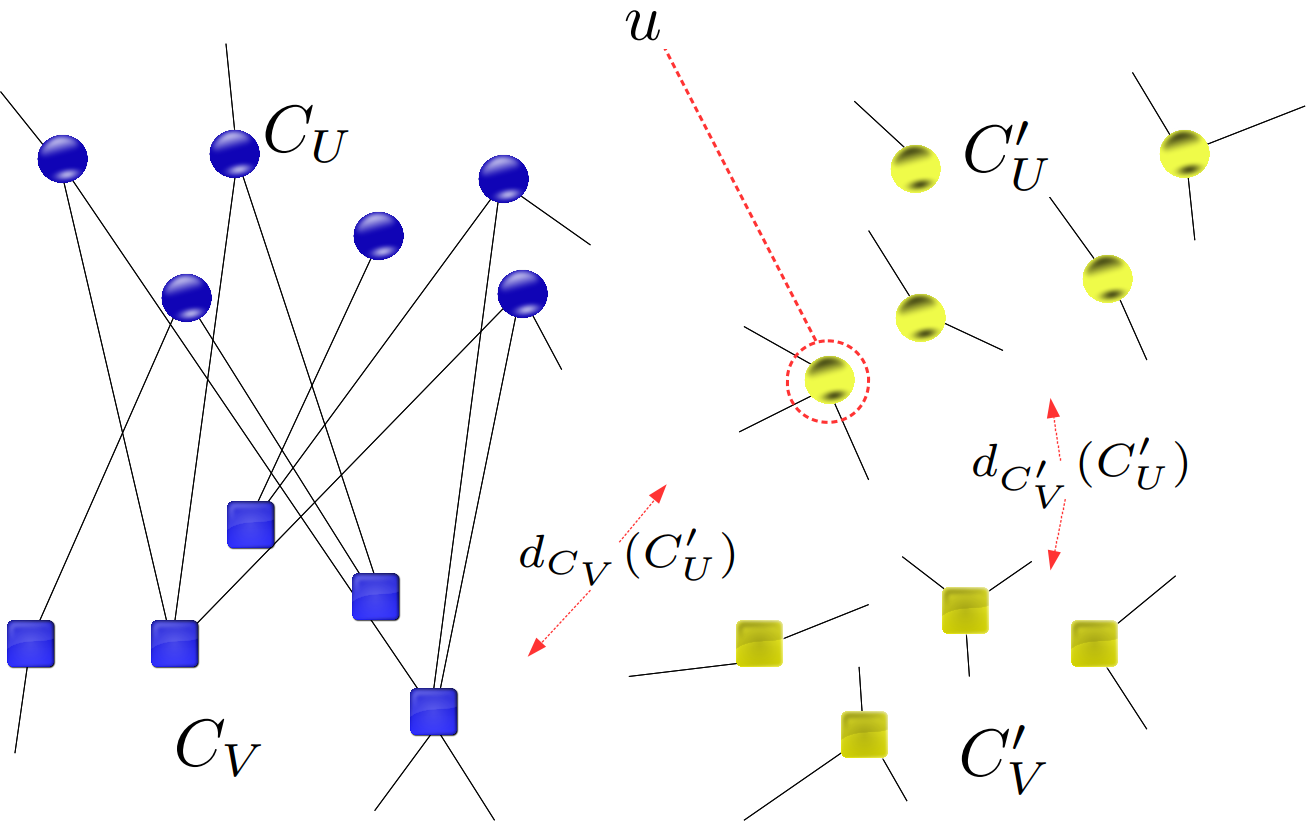}
\caption{\label{fig:bipartite-null} Illustration of bipartite null model for FOCS score. Circles and squares represent $U$ nodes and $V$ nodes, respectively. Blue nodes are the community to be scored. The red circle indicates an arbitrary node $u$ which will have its edges re-assigned under the null.}
\end{figure}

The unipartite null model and the FOCS algorithm can be naturally extended to bipartite networks. The node set of a bipartite network is divided into two sides $U$ and $V$ such that each $u\in U$ can form edges only with nodes in $V$, and vice versa. Consider a candidate bipartite community $C = (C_U, C_V)$, and an exterior node $u\in C_U' := U\setminus C_U$. In the bipartite null, analogously to the unipartite model, all outgoing edges from $C$ and all edges between $C_U'$ and $C_V' := V\setminus C_V$ are broken, and edge stubs coming from $u$ are re-assigned without replacement. In this setting, the degree of $u$ in $C_V$ has the hypergeometric pmf
\begin{equation}\label{eq:bipartite-null}
P(d_u(C_V) = x) \propto \frac{{d_{C_V}(C_U')\choose x}{d_{C_V'}(C_U')\choose d_u - x}}{{d_{C_U'}\choose d_u}}
\end{equation}
The edge-breaking bipartite null model which produces the above distribution is illustrated in Figure 
\ref{fig:bipartite-null}. Using \eqref{eq:bipartite-null} instead of \eqref{eq:basic-null}, the rest of the FOCS approach follows unchanged, with order statistic quantiles computed with respect to the union $C_U'\cup C_V'$. As for directed networks, the use of FOCS depends on the type of community optimization, that is, whether in-degree, out-degree, or joint in-out-degree communities are being optimized. Each of these cases result in similar pmfs to that for undirected unipartite and bipartite cases, and can be used straightforwardly within the general iterative algorithm given above.

\section{Simulations}\label{s:sims}
This section presents simulation results which compare the significance scores of FOCS and existing methods, on communities from both null networks and networks with communities. In all cases, the QS-Test and B-Score methods were run with default parameter settings (as presented in the associated papers and code manuals), and FOCS was run with $p=0.25$.

\subsection{Null Networks}\label{ss:null-networks}
The first simulation experiment involves networks distributed according to the configuration model. Each network had $100$ nodes, and the degree distribution was generated by a power law with exponent $-2$ on the range $[10, 50]$. The total number of simulation repetitions was $1,000$. At each repetition, the Louvain algorithm for modularity maximization was run \citep{blondel2008fast}, and a community for scoring was chosen uniformly at random from the communities in the partition containing more than two nodes.\\
\begin{figure}[!htb]
\centering
\includegraphics[scale=0.5]{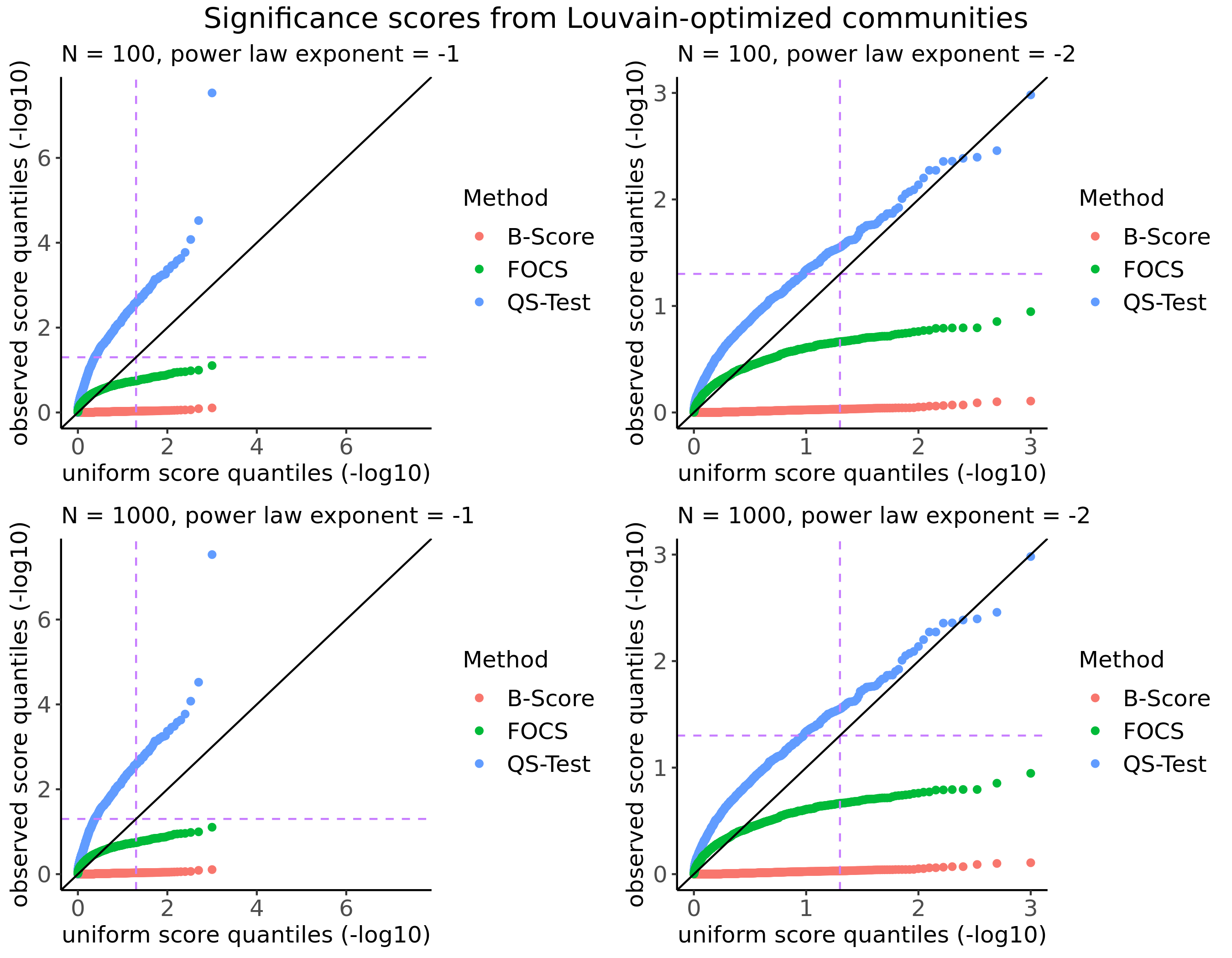}
\caption{\label{fig:null-sims}Significance score distribution on configuration model networks. Purple dotted lines show significance cut-offs.}
\end{figure}

Figure \ref{fig:null-sims} shows the $-\log_{10}$-scale distribution of significance scores from the three methods, plotted against the grid of uniform quantiles that would be expected in a perfectly null distribution of scores. Purple dotted lines show the standard 0.05 significance cutoff on the $\log_{10}$ scale. Therefore, the top-left quadrant formed by the purple dotted lines is the region in which observed scores would indicate significance but uniform-generated scores would not. The bottom-right quadrant is vice-versa. The figure suggests that the QS-Test is anti-conservative on null networks. In other words, applying the QS-Test with a significance cut-off of $\alpha = 0.05$ to a given community will yield a probability of false positive greater than $\alpha$. An explanation for this behavior is not obvious, as the method is performing simulations directly from a null model. The error may be due to poor interaction of the quality function's kernel density estimator with the null model. In contrast, the FOCS and B-Score methods are conservative for $\alpha \leq 0.05$.

\subsection{LFR Networks}\label{ss:lfr-networks}
\begin{figure}[!h]
\centering
\includegraphics[scale=0.6]{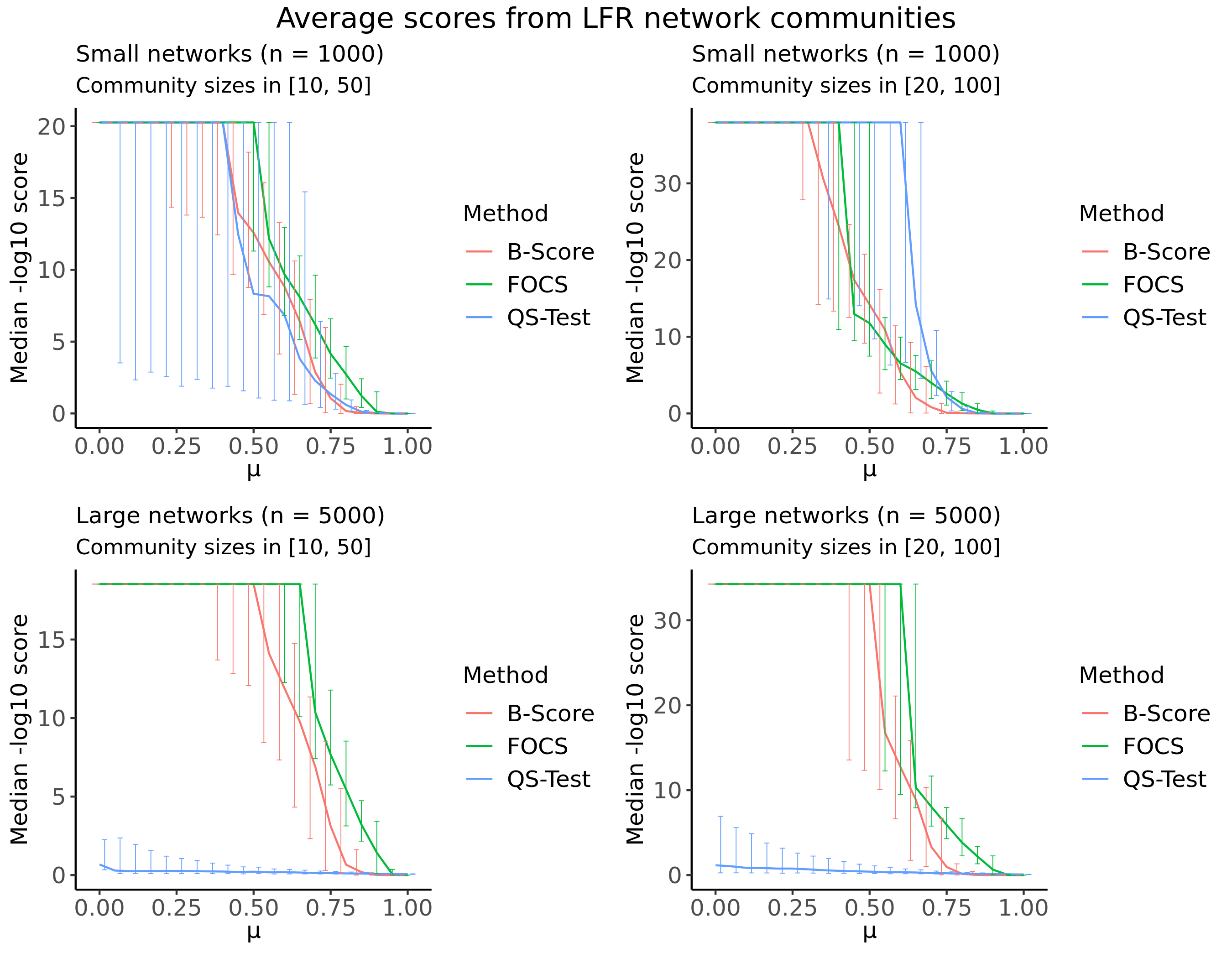}
\caption{\label{fig:lfr-sims} Results from the three methods on the four tested LFR settings. Flat lines are across regions where raw scores went below machine precision.}
\end{figure}

The second simulation experiment involves community-laden networks generated by the LFR  model \citep{lancichinetti2008benchmark}, which will help assess the detection power of each method. The central parameter of this model is $\mu\in[0,1]$, which controls the average proportion of out-edges of each community. If $\mu$ is 1, all edges from each node point outside the node's community, and if $\mu$ is 0, all communities are externally disconnected. Other parameters of the model control the distribution of community sizes and the degree distribution. In this experiment, four LFR network settings are tested: ``small'' networks with $n = 1,000$ vs.\ ``large'' networks with $n = 5,000$, and ``small'' communities with sizes in $[10, 50]$ vs.\ ``large'' communities with sizes in $[20, 100]$. Note that all these networks are tiny by today's industry standard, but that QS-Test and B-Score are prohibitively slow on networks beyond this order of magnitude. In each setting, five LFR networks were simulated at each $\mu$ on an even grid, and the average significance scores for each method were computed across the ground-truth communities from all five repetitions. These average curves, plotted on the $-\log_{10}$ scale so that larger values imply greater significance, are displayed in Figure \ref{fig:lfr-sims}.\\

Figure \ref{fig:lfr-sims} shows that the detection power of the methods vary with both network size and community size. On small networks with small communities, FOCS is the dominant method. On small networks with large communities, FOCS is comparable to B-score, while QS-Test outperforms both these approaches. On large networks, FOCS is the dominant method, and surprisingly, QS-Test loses much of its detection power.


\section{Performance on standard real-world datasets}\label{s:data}
This section presents results from FOCS, B-Score, and QS-Test on real-world datasets commonly used in the networks literature. The datasets used were obtained from the open-access data repository KONECT \citep{kunegis2013konect} and through links provided at Dr.\ Mark Newman's website (\url{http://www-personal.umich.edu/~mejn/netdata/}), and were chosen so that these results could be compared to those from \citep{kojaku2018generalised}. The data sets are listed and briefly described in Table \ref{tab:dataset-desc}, and Table \ref{fig:data-info} provides some of their quantitative characteristics.

\begin{table}[!htb]
\begin{tabular}{p{0.25\textwidth}p{0.75\textwidth}}
\hline\hline
{\bf Network name} & {\bf Description}\\\hline
zachary \citep{zachary1977information} & social ties between karate club members\\
dolphins \citep{lusseau2003bottlenose} & interaction ties between dolphins\\
moreno\_lesmis \citep{knuth1993stanford} & character co-appearance network from Les Miserables\\
enron \citep{shetty2005discovering} & email network from ENRON data\\
netscience \citep{newman2006finding} & collaboration ties between graph researchers \\
polblogs \citep{adamic2005political} & hyperlinks between internet political blogs.\\
airports \citep{kunegis2013konect} & flight network between U.S.\ airports\\
moreno\_propro \citep{jeong2001lethality} & protein interaction network in yeast\\
chess \citep{kunegis2013konect} & chess player game network\\
astro-ph \citep{leskovec2007graph} & collaboration ties between physics researchers\\
internet \citep{kunegis2013konect} & autonomous systems connections network\\\hline\hline
\end{tabular}
\caption{\label{tab:dataset-desc}Description of real-world benchmark datasets.}
\end{table}

\subsection{Detection rates on real data}\label{ss:real-data-sig}
To compare the methods (FOCS, QS-Test, and B-Score) on a particular data set, first the Louvain algorithm was run on the network. Each method was run on each community in the resulting partition, and the proportion of communities with a significance score below 0.05 is shown in Table \ref{fig:data-info}. Two patterns from the simulation study are reflected in these results. FOCS detection rates are more correlated with those from B-Score than those from QS-Test. Second, QS-Test detection rates are much lower on large networks, with the exception of the internet data set, which may be due to the fact that that network had relatively larger communities. These observations suggest that on real data, the methods perform similarly to the simulation study. Note that in this experiment, higher detection rates does not necessarily suggest better performance. For instance, the FOCS method declared two communities significant on the political blogs data set, whereas B-Score declared four. However, the two communities FOCS found significant were the large communities corresponding to (respectively) liberal and conservative sentiments. Other smaller, less-focused communities were ignored, which is a reasonable result.

\begin{table}[ht]
\centering
\begin{tabular}{lllllll}
  \hline\hline
 {\bf Dataset} & {\bf \# Nodes} & {\bf \# Edges} & {\bf \# Comms} & {\bf FOCS} & {\bf B-Score} & {\bf QS-Test} \\ 
  \hline
zachary & 34 & 78 & 4 & 0.250 & 0.500 & 0.500 \\
  dolphins & 62 & 158 & 4 & 0.000 & 0.250 & 1.000 \\
  moreno\_lesmis & 77 & 254 & 6 & 0.167 & 0.333 & 1.000 \\
  enron & 87273 & 1148071 & 378 & 0.889 & 0.958 & 0.423 \\
  netscience & 1589 & 2742 & 177 & 0.819 & 0.887 & 0.525 \\
  polblogs & 1490 & 19090 & 7 & 0.286 & 0.571 & 0.000 \\
  airports & 7976 & 30501 & 23 & 0.957 & 0.739 & 0.522 \\
  moreno\_propro & 1870 & 2277 & 76 & 0.066 & 0.184 & 0.461 \\
  chess & 7301 & 65052 & 36 & 0.722 & 0.583 & 0.361 \\
  astro-ph & 18771 & 198050 & 184 & 0.989 & 0.859 & 0.141 \\
  internet & 34761 & 171402 & 39 & 0.051 & 0.179 & 0.692 \\ \hline\hline
\end{tabular}
\caption{\label{fig:data-info} Summary numbers on the considered data sets: number of nodes, number of edges, and number of communities found by the Louvain algorithm, and proportion of communities found significant (score $ < 0.05$) by each method.}
\end{table}

\subsection{Stability and runtime}\label{ss:stability}
\begin{figure}[!htb]
\centering
\includegraphics[scale=0.4]{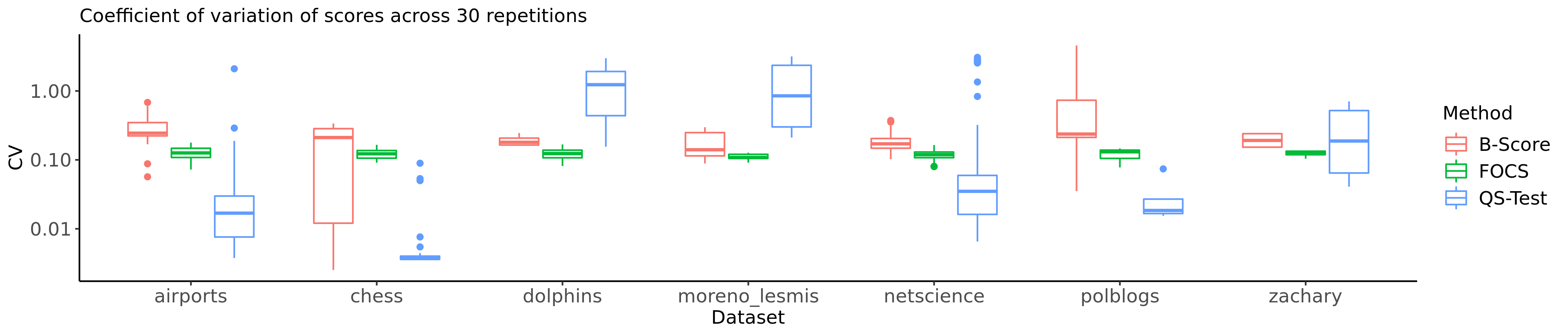}
\caption{\label{fig:pval-cvs} Boxplots of score coefficient of variations across communities, by method and dataset.}
\end{figure}

On some representative small-to-medium-sized real data sets, each method's significance score computation was repeated 30 times with different seeds, for the purposes of measuring (i) numerical stability and (ii) runtime. The larger data sets were not included in this study, as the runtimes for QS-Test and B-Score on these data sets were prohibitively slow. Numerical stability was measured because each method (including FOCS, as described in Section \ref{s:method}) has randomized steps in its algorithm. The metric used to measure stability on a fixed community and network is the coefficient of variation of the significance score, across multiple runs of the algorithm. A low CV score implies that the randomized parts of the method being tested did not drastically affect the significance scores, on the particular community. Figure \ref{fig:pval-cvs} shows the distribution of CV scores (via boxplots) across communities, within each data set. The results show that each method had dominant numerical stability in some data set. However, interestingly, the FOCS CV metrics were by far the most consistent, which suggests that, in contrast to other methods, the expected numerical stability of FOCS scores does not depend on the particular community nor the particular data set, which is desirable. 

The stability and runtime analyses were performed on 2.20 GHz Intel(R) Xeon(R) CPU E7-8890 processors, and the QS-Test computations were distributed across 24 processors, using parallelization options provided with the authors' package (see \url{https://github.com/skojaku/qstest}). Computations for B-Score and FOCS methods were not parallelized. Table \ref{fig:runtimes} gives the mean and standard deviation of runtimes of each method, over the computation repetitions. Note that each runtime (out of thirty runtimes) is the sum of the runtimes from each individual community. On all data sets, FOCS achieved the lowest average runtime compared with the other methods, often by two or three orders of magnitude.

\begin{table}[!htb]
\centering
\tiny
\begin{tabular}{llllllll}
  \hline\hline
Method & airports & chess & dolphins & moreno\_lesmis & netscience & polblogs & zachary \\ 
  \hline
B-Score & 166.54 $\pm$ 3.03 & 1227.21 $\pm$ 72.77 & 0.34 $\pm$ 0.00 & 0.40 $\pm$ 0.00 & 4.92 $\pm$ 0.03 & 158.97 $\pm$ 2.48 & 0.14 $\pm$ 0.08 \\ 
  FOCS & {\bf 1.44} $\pm$ 0.06 & {\bf 5.52} $\pm$ 0.15 & {\bf 0.17} $\pm$ 0.00 & {\bf 0.17} $\pm$ 0.00 & {\bf 0.57} $\pm$ 0.04 & {\bf 0.90} $\pm$ 0.06 & {\bf 0.17} $\pm$ 0.03 \\ 
  QS-Test & 72.04 $\pm$ 0.68 & 348.10 $\pm$ 3.00 & 1.15 $\pm$ 0.09 & 1.42 $\pm$ 0.03 & 26.05 $\pm$ 0.43 & 59.76 $\pm$ 2.28 & 0.81 $\pm$ 0.02 \\
   \hline\hline
\end{tabular}
\caption{\label{fig:runtimes} Average runtime in seconds of methods across 30 repetitions. QS-Test computations were distributed across 24 machines.}
\end{table}


\section{Application to IMDB Data}\label{s:imdb}
This section describes an application of FOCS to a regularly updated IMDB database (\url{https://datasets.imdbws.com}). To display FOCS's handling of diverse network types, a bipartite actor-movie network was constructed from the data. The existing community scoring methods discussed in this paper were not included in this application, because they do not handle bipartite graphs. The movie set was restricted to those released in the US with more than 100 ratings on IMDB, and the actor set was restricted to those with at least one movie from this set. Note that writers and directors were also included as ``actors''. The resulting network had $37,611$ movies, $151,571$ actors, and $362,850$ edges. 

To find optimized communities in the network, we used a simplified, low-cost-exploration simulated annealing algorithm, in the style of the standard method first described in \cite{guimera2005cartography}. Given a community score function and an initial community, the algorithm computes, for each node in the network, the increase in the score possible by moving the node in or out of the community. Ignoring nodes with negative increase, it chooses a node to move with probability proportional to the increase. The algorithm terminates after no score increases are possible, or after a pre-specified number of iterations. The community score function we use is related to bipartite modularity \citep{barber2007modularity}:

\begin{equation}\label{eq:bipartite-mod}
\tfrac{1}{\sqrt{|C_{actors}||C_{movies}|}}\sum_{u\in C_{actors}}\sum_{v\in C_{movies}} A_{uv} - d_ud_v/m
\end{equation}
where $m:=\sum_{u\in V_{actors}}d_u = \sum_{v\in V_{movies}}d_v$. The square-root scaling ensures that trivial increases in the score are ignored; further theoretical motivation for this type of scaling is discussed in \cite{wilson2017community}.

The algorithm described above produced 29,223 communities in the filtered IMDB bipartite network. Each community was scored with FOCS and ranked by decreasing $-\log_{10}$ FOCS score. The extracted bipartite communities with the highest log-scores were those with movie sets that had persistent involvement from all actors in the actor set. Therefore, many of the top-ranked communities featured well-known movie series or collections and their directors, lead writers, and lead actors. Since the null model used by FOCS involves global re-assignment of edge stubs, it makes sense that focused, persistent activity by groups of actors across related films would receive the highest significance scores. In other words, a movie series with a consistent cast should seem most anomalous with respect to the conditional configuration model. A sample of some of the top-ranked communities are shown in Table \ref{tab:imdb-comms}.

\begin{table}[!htb]
\begin{tabular}{|p{0.07\textwidth}|p{0.18\textwidth}|p{0.25\textwidth}|p{0.50\textwidth}|}
\hline\hline
{\bf Rank} & $-\log_{10}(\text{FOCS})$ & {\bf Movie Set} & {\bf Actor Set} \\\hline
1 & 14.75 & {\small 25 out of 28 original 1930s/40s Blondie titles }& {\small Chic Young, Arthur Lake, Frank R. Strayer, Larry Simms, Marjorie Ann Mutchie, Penny Singleton}\\\hline
2 & 14.60 & {\small Super Troopers, Super Troopers 2, Beerfest, The Slammin' Salmon, Club Dread} & {\small Kevin Hefferman, Nathan Barr, Erik Stolhanske, Jay Chandrasekhar, Richard Perello, Steve Lemme, Paul Soter}\\\hline
6 & 13.69 & {\small All seven major Harry Potter films} & {\small Daniel Radcliffe, Steve Kloves, J.K. Rowling, Emma Watson, David Yates, Lorne Orleans, David Barron, Rupert Grint, David Heyman}\\\hline
22 & 11.01 & {\small All five major Twilight films (Twilight and four Twilight Saga titles)} & {\small Stephenie Meyer, Karen Rosenfelt, Melissa Rosenberg, Kristen Stewart, Wyck Godfrey, Taylor Lautner, Robert Pattinson}\\\hline\hline
\end{tabular}
\caption{\label{tab:imdb-comms} Some top-ranked IMDB bipartite communities with well-known titles, ordered by FOCS score. Omitted titles exhibited the same persistence of movie series theme and actor participation.}
\end{table}

\begin{figure}[!htb]
\centering
\includegraphics[scale=0.4]{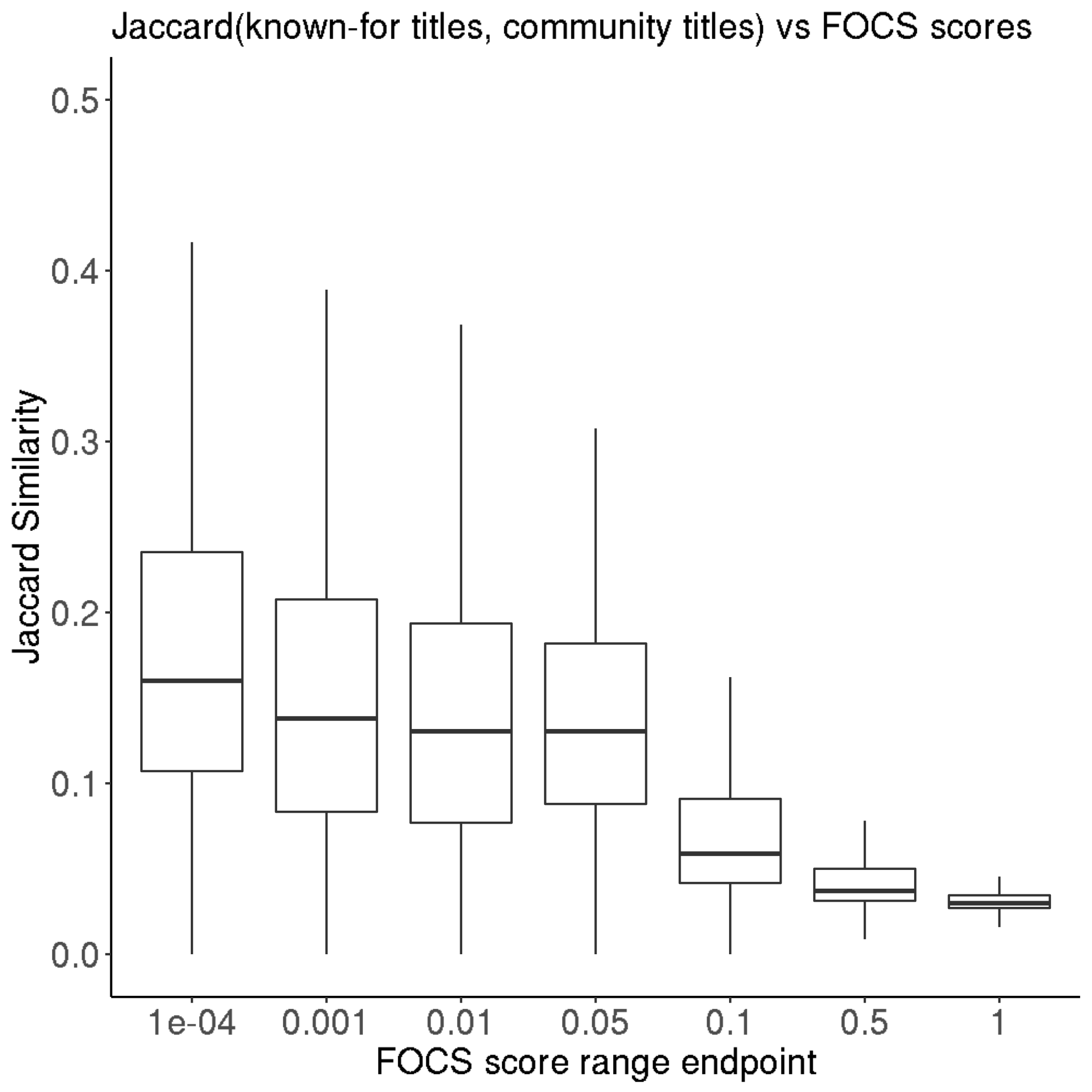}
\caption{\label{fig:imdb-score} Distribution of jaccard similarities between cluster movie sets and actor ``known-for"  movie sets, across clusters, within ranges of FOCS scores. $x$-axis labels display the upper endpoint of the range, which extends back to the previous (left) upper endpoint. The lowest range extends to zero.}
\end{figure}

To quantitatively assess the true intra-relatedness of each community, the jaccard similarity between the movie set and the union of the sets of movies that each actor is ``known for'', according to the IMDB metadata, was computed. This score correlated highly with FOCS scores. In particular, the median jaccard similarities were increasingly large as ranges of the FOCS scores decreased on a quasi-logarithmic scale (see Figure \ref{fig:imdb-score}). This shows that, in this application, the community ranking and threshold provided by FOCS aligned with ground-truth signal.

Interestingly, the majority of communities produced by the node-swapping algorithm were not significant. Communities with large FOCS scores exhibited much less internal coherence - many communities contained mostly unrelated movies with few actor overlaps. Simply put, these communities were poor local maxima of the algorithm. It is a particular utility of a method like FOCS to be able to distinguish between these communities and meaningful, strongly-connected communities.

\section{Discussion}\label{s:discussion}
This paper introduces new models and tests for optimized communities in networks, and presents FOCS, a new algorithm for significance scoring that has performance benefits over of existing approaches. FOCS uses a core scoring approach that exploits the fact that communities are rarely optimized perfectly, and therefore weakly connected nodes in communities distribute edges approximately according to a random graph null model. Because of this, FOCS has a simplicity that previous methods lack, making it more scalable, more numerically stable, and more generalizable. Despite its simplicity and speed, FOCS performs ahead of or comparably to preceding methods in terms of reduced tendency for false positives, and reduced significance scores on true communities. On a large-scale bipartite movie-actor network derived from IMDB data, the highest FOCS-ranked communities produced by an extraction method were those with highly related movie sets sharing continuous involvement from a dedicated cast and crew. This suggests that FOCS can be useful in detecting communities exhibiting anomalous, persistent involvement from its constituents. 

The FOCS method has some limitations. First, as with the existing methods, FOCS uses resampling methods in its computation. Additionally, FOCS is based on arguably plausible yet non-rigorous ideas about the distribution of nodes in optimized communities. Therefore, FOCS is not an exact statistical test, and its results should be reported with these caveats. It should be noted that existing methods also rely on approximations, which is often necessary when dealing with the intractable distributions presented by graph models. Finally, the simulations on null networks in Section \ref{ss:null-networks} showed that FOCS may be overly-conservative. This means that there may be headroom to improve FOCS by making it less conservative and more powerful, which is an area for future research. 

Despite these limitations, the FOCS method appears to improve greatly on the existing options for scoring the significance of individual communities. Given its scalability and straightforward implementation, it can be readily used in real-time anomaly detection, machine learning pipelines, and scientific studies. The basic implementation of the FOCS method used in experiments discussed in this paper can be found at \url{https://github.com/google/fast-optimized-community-significance}, and the pipeline of experiments can be reproduced with code at \url{https://github.com/jpalowitch/focs\_experiments}.

\begin{acknowledgments}
The author acknowledges helpful conversations with Andrea Lancichinetti and Peter J.\ Mucha.
\end{acknowledgments}

\bibliography{focs_aps}

\end{document}